\documentclass{article}

\usepackage{psfrag,epsfig,graphicx,graphics,xcolor}
\usepackage{wrapfig}
\usepackage{amsmath}
\usepackage{amssymb}
\newcommand{\bfr}{\begin{flushright}}
\newcommand{\efr}{\end{flushright}}
 
\begin{document}
\title{High energy rho meson leptoproduction 
\thanks{Presented at  the Low x workshop, May 30 - June 4 2013, Rehovot and
Eilat, Israel}
}

\author{
Adrien Besse$^1$, Lech Szymanowski$^2$, Samuel Wallon$^{3,4}$\\
{\small $^1$Irfu-SPhN, CEA, Saclay, France}\\
{\small $^2$National Centre for Nuclear Research (NCBJ), Warsaw, Poland}\\
{\small $^3$LPT, Universit\'e Paris-Sud, CNRS, 91405, Orsay, France}\\
{\small $^4$UPMC Univ. Paris 06, Facult\'e de physique,}\\
{\small 4 place Jussieu, 75252 Paris Cedex 05,
France
}
\smallskip\\
}

\date{\today
}

\maketitle
\begin{abstract}
We investigate the longitudinal and transverse polarized cross-sections of the leptoproduction of the $\rho$ meson in the high energy limit. Our model is based on the computation of the impact factor $\gamma^*(\lambda_{\gamma})\to\rho(\lambda_{\rho})$ using the twist expansion in the forward limit and expressed in the impact parameter space. This treatment involves in the final stage the twist~2 and twist~3 distribution amplitudes (DAs) of the $\rho$ meson and the dipole scattering amplitude. Taking models that exist for the DAs and for the dipole cross-section, we get a phenomenological model for the helicity amplitudes, we compare our predictions with HERA data and get a fairly good description for large enough virtualities of the photon.  
\\
~
\\
PACS number(s): 13.60.Le, 12.39.St, 12.38.Bx
\end{abstract}

\newcommand{\kb}{\underline{k}}
\newcommand{\garho}{\gamma^*_{\lambda_{\gamma}}\to \rho_{\lambda_{\rho}}}
\newcommand{\garhoL}{\gamma^*_{L}\to \rho_{L}}
\newcommand{\garhoT}{\gamma^*_{T}\to \rho_{T}}
\newcommand{\qb}{\bar{q}}
\newcommand{\rb}{\underline{r}}
\newcommand{\yb}{\bar{y}}
\newcommand{\nn}{\nonumber}

\section{Introduction}
We study the high energy diffractive leptoproduction of $\rho$ meson
\begin{equation}
\label{DVMP}
\gamma^*(q,\lambda_{\gamma})\,N(p)\to \rho(p_{\rho},\lambda_{\rho})\,N(p')\,,
\end{equation}
where $N$ is the nucleon target, $\lambda_{\rho}$ and $\lambda_{\gamma}$ are respectively the polarizations of the $\rho$ meson and of the virtual photon. 
The longitudinal and transverse polarized cross-sections $\sigma_L$ and $\sigma_T$ of the process (\ref{DVMP}) can be expressed in terms of the helicity amplitudes, which are denoted $T_{\lambda_{\rho}\lambda_{\gamma}}$. In the limit of high energy in the center of mass of the $\gamma^* \,N$ system, the helicity amplitudes can be factorized, using the $k_T-$factorization scheme, into the convolution of the impact factor $\Phi^{\gamma^*_{\lambda_{\gamma}}\to\rho_{\lambda_{\rho}}}$ associated to the process
\begin{equation}
\gamma^*(q,\lambda_{\gamma})\,g(k_1)\to\rho(p_{\rho},\lambda_{\rho})\,g(k_2)\,,
\end{equation}
and the unintegrated gluon density\footnote{We denote by $\underline{x}$ the 2-dimension euclidean vector associated to the Minkowskian $x_{\perp}$, $\underline{x}^2=-x_{\perp}^2$.} $\mathcal{F}(x,\kb)$. In our kinematics we use the Sudakov decomposition along the light cone vectors $p_1$ and $p_2$, such as
\begin{equation}
p_{\rho}\sim p_1\,,\quad p\sim p_2 \,, \quad q\sim p_1- \frac{Q^2}{s}p_2\,, \quad s=(q+p)^2\sim 2 p_1\cdot p_2\gg (Q^2\,,\,\,m_{\rho}^2)\,.
\end{equation}
\begin{figure}[htbp]
	\centering
\psfrag{P}[cc][cc]{$p$}
\psfrag{P'}[cc][cc]{$p'$}
\psfrag{Gam}[cc][cc]{$\gamma^*$}
\psfrag{Rho}[cc][cc]{$\rho$}
\psfrag{k}[cc][cc]{$k_1$}
\psfrag{k2}[cc][cc]{$k_2$}
\psfrag{Phi}[cc][cc]{${\Phi^{\gamma^*_{\lambda_{\gamma}}\to\rho_{\lambda_{\rho}}}}$}
\psfrag{Phibis}[cc][cc]{$\mathcal{F}(x,\kb)$}
\includegraphics[width=0.5\linewidth]{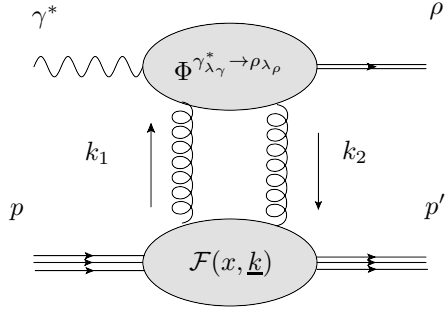}
\caption{Impact factor representation of the helicity amplitudes.}
	\label{ImpFact}
\end{figure}
The $t$-channel gluon momenta, illustrated in fig.~\ref{ImpFact}, read $k_1=\frac{\kappa+Q^2+\kb^2}{s}p_2+k_{\perp}$ and $k_2=\frac{\kappa+\kb^2}{s}p_2+k_{\perp}$, where $\kappa$ is the energy in the center of mass of the system $\gamma^*(q)\, g(k_1)$. The helicity amplitudes are 
\begin{equation}
\label{ConvImp}
T_{\lambda_{\rho}\lambda_{\gamma}}= is \int \frac{d^2\kb}{(\kb^2)^2} \, \Phi^{\gamma^*_{\lambda_{\gamma}}\to\rho_{\lambda_{\rho}}}(\kb) \, \mathcal{F}(x,\kb)\,.
\end{equation}

Assuming the virtuality of the photon $Q$ ($Q^2=-q^2$) is large compared to the QCD scale $\Lambda_{QCD}$, the impact factors $\Phi^{\gamma^*_{L}\to\rho_{L}}$ and $\Phi^{\gamma^*_{T}\to\rho_{L}}$ 
 were computed in ref.~\cite{GinzburgPanfilSerbo}, using the collinear factorization on the light-cone. In this approach, the impact factors are parameterized by the leading twist DA of the $\rho$ meson. 
 This computation was extended in refs.~\cite{Anikin2009,Anikin2010} to obtain the $\Phi^{\gamma^*_{T}\to\rho_{T}}$ impact factor in the limit $\left|t\right|\sim 0$. In this last case, the twist~2 contribution vanishes and the amplitude is parameterized by the twist~3 DAs of the $\rho$ meson. The result for $\Phi^{\gamma^*_T\to\rho_T}$ obtained from the light-cone collinear factorization is the sum of two contributions: from a quark antiquark ($q\qb$) Fock state and from a quark antiquark gluon ($q\qb g$) Fock state. 
Relations between the DAs can be derived from the first principles of QCD and the twist~3 DAs that parameterize the Fourier transforms of the  $q\qb$ correlators can be split into two solutions: the Wandzura-Wilczek (WW) solutions, which consist in neglecting the $q\qb g$ DAs, and the "genuine" solutions, that only depend on the $q\qb g$ DAs. Thus, one can represent the $q\qb$ and the $q\qb g$ contributions to the impact factor $\Phi^{\gamma^*_T\to\rho_T}$ as a sum of a WW contribution and of a genuine contribution. A first phenomenological model proposed in ref.~\cite{Anikin2011} was based on the results of refs.~\cite{GinzburgPanfilSerbo, Anikin2010} and used a model for the proton impact factor inspired from ref.~\cite{GunionSoper}. The results of this study have led to the conclusion that the soft $t-$channel gluons have a sizable contribution, which calls for the implementation of the saturation effects in this perturbative approach.

 For this aim, in ref.~\cite{Besse2012}, we have performed calculations of the twist~2 and twist~3 impact factors in the impact parameter space. We have shown also the equivalence of obtained results with the ones in momentum space of ref.~\cite{Anikin2010}. The results in the impact parameter representation can be put in the form 
\begin{eqnarray}
\label{phiLpsi}
\Phi^{\gamma^{*}_L \rightarrow \rho_L}(\kb,Q,\mu^2)&=&\left(\frac{\delta^{ab}}{2}\right)\int dy \int d \rb\,\, \psi^{\gamma^*_L\to\rho_L}_{(q\qb)}(y,\rb;Q,\mu^2)\, \mathcal{A}(\rb,\kb)\,,\\
\Phi^{\gamma^{*}_T \rightarrow \rho_T}(\kb,Q,\mu^2)&=&\left(\frac{\delta^{ab}}{2}\right)\int dy \int d \rb\,\, \psi^{\gamma^*_T\to\rho_T}_{(q\qb)}(y,\rb;Q,\mu^2) \,\mathcal{A}(\rb,\kb)\nn\\
&&\hspace{-2cm}+\left(\frac{\delta^{ab}}{2}\right)\int dy_2 \int dy_1 \int d \rb \,\, \psi^{\gamma^*_T\to\rho_T}_{(q\qb g)}(y_1,y_2,\rb;Q,\mu^2) \mathcal{A}(\rb,\kb)\,,
\label{phiTpsi}
\end{eqnarray}
where the functions  $\psi^{\gamma^*_L\to\rho_L}_{q\qb}$, $\psi^{\gamma^*_T\to\rho_T}_{q\qb}$ and $\psi^{\gamma^*_T\to\rho_T}_{q\qb g}$ are respectively our results for the transitions $\gamma^*_L\to(q\qb)\to\rho_L$, $\gamma^*_T\to (q\qb)\to\rho_T$ and $\gamma^*_T\to(q\qb g)\to\rho_T$. $\mathcal{A}(\rb,\kb)$ is the scattering amplitude of a color dipole of transverse size $\rb$, with the $t-$channel gluons having transverse momenta $\kb$. In eqs.~(\ref{phiLpsi}, \ref{phiTpsi}) $a$ and $b$ are the color indices of the $t-$channel gluons in a singlet state. 
 As a result, the well-known wave functions of the virtual photon factorize out in the expressions of $\psi^{\gamma^*_{L}\to\rho_{L}}_{q\qb}$ and $\psi^{\gamma^*_{T}\to\rho_{T}}_{q\qb}$. 
The $\rho$ meson non-perturbative parts are encoded by the twist~2 and twist~3 DAs and $\mu$ stands for the factorization/renormalization scale of the DAs. We use the model of Ball, Braun, Koike and Tanaka developed in ref.~\cite{Ball:1998sk} to get explicit expressions for the DAs. This model relies on the conformal expansion of the DAs to separate the longitudinal momentum dependence from the scale dependence in $\mu$. 
 It is customary to call "asymptotic" (AS) the results in the limit $\mu^2\to\infty$. On the other hand, a natural choice for this scale is $\mu^2=(Q^2+m_{\rho}^2)/4$. 
Note that the factorization of the dipole scattering amplitude $\mathcal{A}(\rb,\kb)$ is due to the relations between the DAs coming from the equations of motion of QCD.

Inserting the expressions (\ref{phiLpsi}, \ref{phiTpsi}) for the impact factor in eq.~(\ref{ConvImp}) leads to
\begin{eqnarray}
\label{T00Lpsi}
\frac{T_{00}}{s}\!&\!=\!&\!\int dy \int d \rb\, \psi^{\gamma^*_L\to\rho_L}_{(q\qb)}(y,\rb;Q,\mu^2) \,\hat{\sigma}(x,\rb)\,,\\
\label{T11Tpsi}\frac{T_{11}}{s}\!&\!=\!&\! \int d \rb\left[\int \! dy\, \psi^{\gamma^*_T\to\rho_T}_{(q\qb)}(y,\rb;Q,\mu^2)
\right.\\&&\hspace{-1cm}
\left.+\int \! dy_2 \int \! dy_1  \psi^{\gamma^*_T\to\rho_T}_{(q\qb g)}(y_1,y_2,\rb;Q,\mu^2)\right] \hat{\sigma}(x,\rb)\,,\nn
\end{eqnarray}
where $\hat{\sigma}(x,\rb)$ is the dipole cross-section. These expressions are the starting point for our phenomenological analysis.

\begin{figure}[h]
	\centering
	\begin{tabular}{c}
		\includegraphics[width=0.7\textwidth]{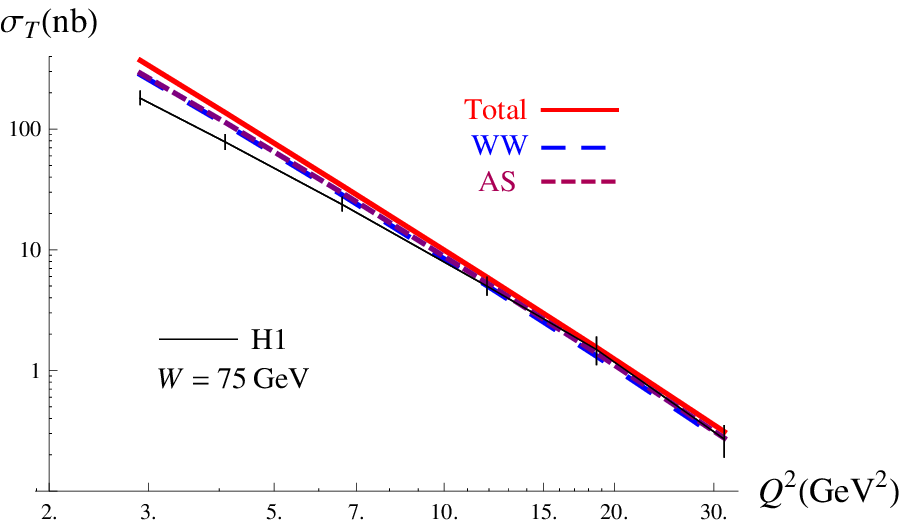}\\
		\includegraphics[width=0.7\textwidth]{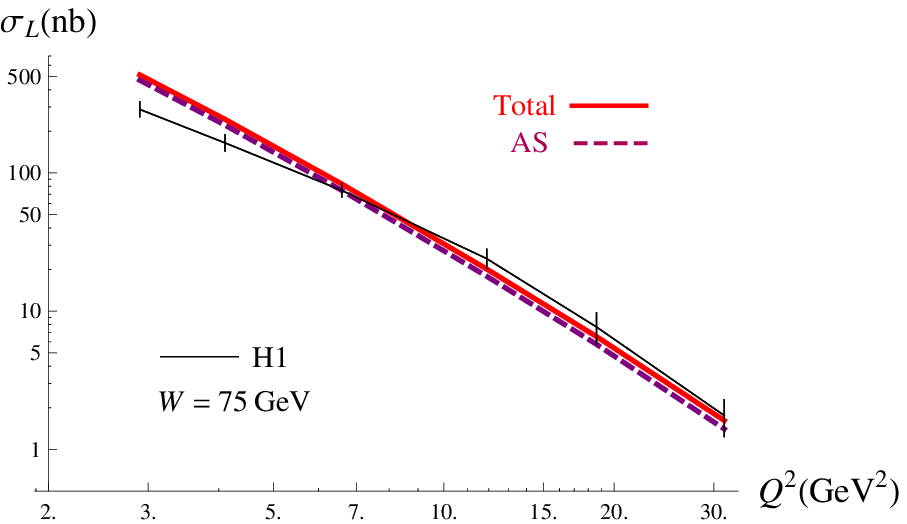}
	\end{tabular}
	\caption{Left: Total, WW and AS contributions to $\sigma_T$ vs $Q^2$, compared to H1 \cite{H1} data. Right: Total and AS twist~2 contributions to $\sigma_L$ vs $Q^2$ compared to H1 data. }	\label{Fig3}
\end{figure}	
	
\section{Confronting our predictions with HERA data}

In ref.~\cite{Besse:2013muy}, we have compared our predictions for the transverse and longitudinal polarized cross-sections, shown in fig.~\ref{Fig3}, with the data from H1 \cite{H1}. These predictions are obtained using the dipole scattering amplitude of ref.~\cite{Albacete:2010sy}, which is based on numerical solutions of the running coupling  Balitsky-Kovchegov (rcBK) equation \cite{Balitsky:2006wa}. This model of dipole scattering amplitude allows to account for the saturation effects in our description of the $\rho$ meson leptoproduction. Note that as we use a model of dipole cross-section already fitted on inclusive structure functions then we do not need to adjust value of any parameter. 
 The results are in good agreement with the data for $Q^2 \gtrsim 5\;$GeV$^2$ and they are weakly dependent on the choice of the factorization/renormalization scale $\mu$. The discrepancy for smaller virtualities $Q^2 \lesssim 5\;$GeV$^2$ indicates that higher twist corrections to the impact factors can become important for such values of $Q^2$.

\begin{figure}[h]
	\centering
	\begin{tabular}{c}
		\includegraphics[width=0.7\textwidth]{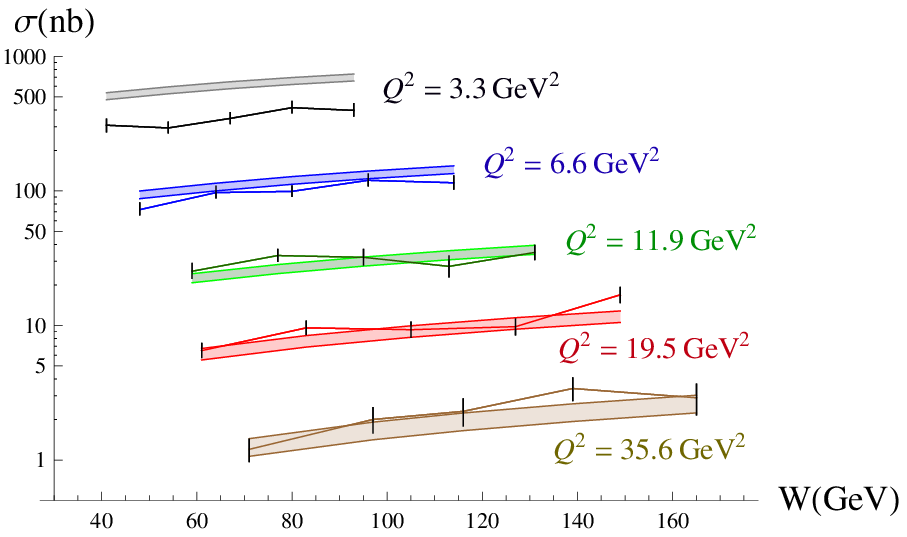}	\\
	\includegraphics[width=0.7\textwidth]{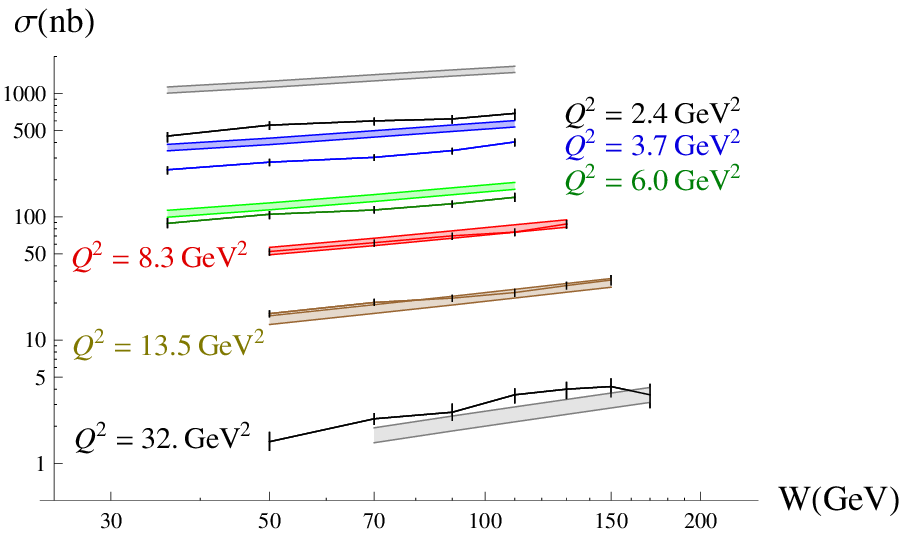}
	\end{tabular}
	\caption{
	Predictions for the total cross-section $\sigma$ vs $W$ compared to H1 \cite{H1} (left) and ZEUS \cite{ZEUS} (right) data.}
	\label{Fig4}
\end{figure}

In fig.~\ref{Fig4}, we show our predictions for the total cross-section $\sigma$ of the diffractive leptoproduction of $\rho$ meson and compared then with the data of H1 \cite{H1} and ZEUS \cite{ZEUS}, as a function $W$. 
The $W-$dependence of our predictions is given by the dipole cross-section model \cite{Albacete:2010sy}. In this way we obtain a good agreement between the predictions and the data for the $W-$dependence.

\section{Conclusions}
The success of the model we have presented to describe the $W-$ and the $Q^2-$dependencies with the proper normalizations for large enough $Q^2$, relies on the computations from first principles of the impact factors $\Phi^{\gamma^*\to\rho}$ and the models for the twist~2 and twist~3 DAs as well as the model for the dipole scattering amplitude. 
Consequently, this approach constitutes a good way to unravel the non-perturbative aspects of the leptoproduction of the $\rho$ meson. 
The perspectives of this study are numerous, as it could be extended in the non-forward kinematics and for other helicity amplitudes. This could allow to probe the impact parameter dipole/nucleon target dependence of the dipole scattering amplitudes. The higher twist correction effects could lead to a better description of the data for lower values of $Q^2$ closer to the saturation scale in the HERA kinematics.

\section{Acknowledgment}
We thank B. Duclou\'e, K. Golec-Biernat, C. Marquet, S. Munier and B. Pire for interesting discussions and comments on this work. 
We thank the organizers, the Tel Aviv French ambassy and the French CEA (IPhT and DSM) for support. This work is supported by the P2IO consortium, the Polish Grant NCN No DEC-2011/01/D/ST2/03915, the Joint Research Activity Study of Strongly Interacting Matter (acronym HadronPhysics3, Grant 283286) under the Seventh Framework Programme of the European Community and the French grant ANR PARTONS (ANR-12-MONU-0008-01).



\bibliographystyle{apsrev4-1}

\begin{thebibliography}{99}
\bibitem{GinzburgPanfilSerbo} I. F. Ginzburg, S. L. Panfil, V. G. Serbo, Nucl. Phys. B {\bf 284}, 685-705 (1987).
\bibitem{Anikin2009} I. V. Anikin, D. Yu. Ivanov, B. Pire, L. Szymanowski, S. Wallon, Phys. Lett. B {\bf 682}, 413-418 (2010).
\bibitem{Anikin2010} I. V. Anikin, D. Yu. Ivanov, B. Pire, L. Szymanowski, S. Wallon, Nucl. Phys. B {\bf 828}, 1-68 (2010).
\bibitem{GunionSoper} J. F. Gunion and D. E. Soper, Phys. Rev. D  {\bf 15}, 2617 (1977).
\bibitem{Anikin2011} I. Anikin, A. Besse, D. Yu. Ivanov, B. Pire, L. Szymanowski, S. Wallon, Phys. Rev. D  {\bf 84}, 054004 (2011).
\bibitem{Besse2012} A. Besse, L. Szymanowski, S. Wallon, Nucl.\ Phys.\  B {\bf 867},  19-60 (2013).
\bibitem{Ball:1998sk}
  P.~Ball, V.~M.~Braun, Y.~Koike and K.~Tanaka,
  Nucl.\ Phys.\ B {\bf 529}, 323 (1998).
\bibitem{Besse:2013muy}  A. Besse, L. Szymanowski, S. Wallon, arXiv:1302.1766 [hep-ph].
\bibitem{H1} F. D. Aaron {\it et. al.} [H1 Collaboration], JHEP  {\bf 1005}, 032 (2010).

\bibitem{Albacete:2010sy}
J.~L. Albacete, N.~Armesto, J.~G. Milhano, P.~Quiroga-Arias, and C.~A. Salgado, Eur. Phys. J. C {\bf 71}, 1705 (2011).
\bibitem{Balitsky:2006wa}
  I.~Balitsky,
  Phys.\ Rev.\ D {\bf 75} (2007) 014001
  [hep-ph/0609105].
\bibitem{ZEUS} S. Chekanov {\it et al.} [ZEUS Collaboration], PMC Phys. A  {\bf 1}, 6 (2007).

\end{thebibliography}

\end{document}